# Effect of tip-geometry on contrast and spatial-resolution of the Near-Field Microwave Microscope

Atif Imtiaz and Steven M. Anlage[a]

Center for Superconductivity Research, Department of Physics, University of Maryland,

College Park, MD 20742-4111

**Abstract**

The Near-Field Scanning Microwave Microscope (NSMM) can quantitatively image materials properties at length scales far shorter than the free space wavelength ($\lambda$). Here we report a study of the effect of tip-geometry on the NSMM signals. This particular NSMM utilizes scanning tunneling microscopy (STM) for distance-following control. We systematically examined many commercially available STM tips, and find them to have a conical structure on the macroscopic scale, with an embedded sphere (of radius $r_{sphere}$) at the apex of the tip. The $r_{sphere}$ values used in the study ranged from 0.1 μm to 12.6 μm. Tips with larger $r_{sphere}$ show good signal contrast (as measured by the frequency shift ($\Delta f$) signal between tunneling height and 2 μm away from the sample) with NSMM. For example, the tips with $r_{sphere}$ = 8 μm give signal contrast of 1000 kHz compared to 85 kHz with a tip of $r_{sphere}$ = 0.55 μm. However, large $r_{sphere}$ tips distort the topographic features acquired through STM. A theoretical model is used to understand the tip-to-sample interaction. The model quantitatively explains the measured change in quality factor (Q) as a function of height over bulk Copper and Silicon samples.

---

[a] email address: anlage@umd.edu

PACS: 78.70.Gq, 42.25.Bs, 41.20.-q, 84.37.+q

# I. Introduction:

Near-field microscopy techniques have been used to break the conventional far-field Abbe's limit in the spatial-resolution of an image[1] formed with electromagnetic waves. Near-field microscopy has been successfully done at optical[2,3] and microwave[4,5] frequencies. In a typical experimental setup, the distance h between the probe and sample is much less than the free space wavelength λ. Even when the condition h << λ is fulfilled, the spatial-resolution can be improved further by bringing the probe closer to the sample (in a non-destructive fashion) to enhance the field localization. The Scanning Tunneling Microscope (STM) has been an important tool in this regard, since it allows the probe (or the tip) to be brought to a nominal height of 1 nanometer above the sample, non-destructively.[6] Here we systematically examine the effect of tip geometry on both the NSMM and STM images in a combined microscope, and present a model consistent with the data.

The previous attempts to integrate the microwave microscope with STM can be put in three broad categories. The first combined microscope[7-12], integrates the STM with near-field microwave transmission measurements. In this case, the sample is uniformly illuminated with a frequency of interest and an antenna (in the near-field) picks up the signal for measurement. The same antenna is used as the STM tip. The major accomplishment here was in understanding the effect of surface topography on the complex transmission coefficient.[12] An experiment was performed on a 7 nm thick Pt/Carbon film on a Si/SiO$_2$ substrate. There was a 2 nm deep depression in the Pt/C film and as the STM scanned across the depression, the frequency shift signal showed the same qualitative response as the topography. To demonstrate high resolution of such a microscope, contrast due to mono-atomic steps in Cu(111) surface were imaged.[10]

The second class of combined microscopes[13-17] integrates the STM with a resonant microwave cavity. The sample is generally inside the resonant cavity and the tip (to perform STM) is brought into the cavity through a hole made on one of the walls of the cavity (the hole dimension is much smaller than the wavelength of incident microwaves). Microwaves are injected locally in to the sample at a frequency that is resonant with the cavity. One can also send in microwave signals at a frequency that is exactly one half or one third of the resonant frequency of the cavity.[16] Harmonics produced locally by the sample will then excite the cavity resonance. A loop antenna is set some where in the cavity where the magnetic field of the resonant mode is maximum to pick up a transmitted or generated signal. Notable accomplishments are studies of different metal surfaces to show high resolution images of the third harmonic signal. In one case self assembled mono-layers (made from mixture of chemicals perflourononanoyl-2-mercahptoethylamide) on a gold surface were studied to show high z-resolution[14] and in another a $WSe_2$ surface was studied to show high spatial-resolution[17] in the third harmonic signal while simultaneously an STM topography image was acquired.

The third class of combined microscopes[18-22] couples an STM tunnel junction with laser light. The tunnel junction is illuminated with two fine-tuned optical frequencies. The non-linear IV characteristic of the tunnel junction is used to detect the rectification signal, the sum and difference frequencies. This technique is used to detect higher harmonics as well. One notable experiment performed with such a setup is to simultaneously acquire the tunneling current and the difference frequency signal ($\Delta\omega$) over a graphite surface.[20] The $\Delta\omega$ signal was also used for distance control over the surface to construct a topography image. Another notable experiment (in which a scanning force microscope (SFM) was used) measured the $\Delta\omega$ signal on a pattern of small metal islands (gold) which was on top of a non-conducting BaF$_2$ substrate.[21] A qualitative map of conductivity was made to distinguish between conducting and non-conducting regions.

The above work is important, however, it lacks the ability to quantitatively extract and understand the materials contrast. In order to overcome this shortcoming, we describe a novel near-field scanning microwave microscope (NSMM) which uses STM for distance-following. The resonator for this NSMM is made up of a transmission line, and quantitative materials contrast with this type of microscope has already been discussed[4,5] in the literature. Our goal with this STM-assisted experiment is quantitative high resolution materials contrast for a similar microscope.

## II. Experiment:

The unique feature of our NSMM[23] (schematic shown in Fig. 1) is the use of a transmission line resonator made out of commercially available UT-085 coaxial cable. The length of the resonator is 1.06m and on one end it terminates at an open ended coaxial probe and on the other end it is connected to a home-made decoupling capacitor (labeled decoupler in Fig. 1). The open ended coaxial probe is just a piece of coaxial cable in which the center conductor is replaced by a stainless steel capillary tube[17] (of hollowed inner diameter 0.005" labeled in the inset of Fig. 1). This capillary tube holds the STM tip, which is also used to perform microwave microscopy. The bias-Tee (of typical bandwidth from 0.04 to 18 GHz) permits the use of the same tip for both microscopes. The inductor of the bias-Tee allows one to connect the STM feedback circuit to the inner conductor of the transmission line resonator. The same inductor damps out the ac (microwave) signal from interfering with the operation of the STM. The capacitor of the bias-Tee stops the dc signal from reaching the microwave circuit and this capacitor is in series with the decoupling capacitor. The value of bias-Tee capacitor is at least an order of magnitude higher than the decoupling capacitor, so the effective capacitance for the microwave signal is that of the decoupling capacitor.

The transmission line resonator is connected via the directional coupler to the microwave source and the diode detector (HP 8473C diode detector with bandwidth 0.01 to 26.5 GHz) and the output from this detector is sent to the two lock-in amplifiers referenced at the external oscillator frequency $f_{mod}$. The microwave source is frequency modulated at a rate of $f_{mod}$ as well, and with the help of a feedback circuit (FFC in Fig. 1), the source is kept locked at the resonant frequency of the resonator.[25] One lock-in amplifier is part of the frequency following circuit[24] (FFC- as shown inside the dashed box). The output of this lock-in amplifier is integrated over time to measure the frequency shift ($\Delta f$) of the transmission line resonator. The lock-in amplifier outside the dashed box in Fig. 1 picks up the signal at twice $f_{mod}$ (labeled as $V_{2f}$ in Fig. 1) which gives a measure of the quality factor[25] (Q) of the resonator. The $\Delta f$ and the $f_{mod}$ signals from the oscillator are added together and sent to the microwave source (to keep it locked at the resonant frequency of the resonator) and this completes the microwave feedback circuit. Over time, we used many different microwave sources, including un-synthesized WaveTek 904, WaveTek 907 sources and later synthesized (for better signal stability) HP83620B and Agilent 8257E sources.

The commercially available STM used was the CryoSXM manufactured by Oxford Instruments. This commercial package included an STM head assembly, electronics, software and a cryostat. The STM head assembly, probe arm, electronics and software combined together are called TOPSystem3 (or TOPS3). The Oxford cryostat has the ability to vary the sample temperature from about 4.2K to 300K. The sample for the experiment sat on the XYZ piezo stage of this STM. With the STM tunnel junction established, the nominal height of the tip above the sample is 1 nm.[6] As a result the STM/NSMM probe scans across the sample non-destructively. There are three independent signals collected from this integrated microscope, which includes the frequency shift ($\Delta f$) and Q of the resonator (due to variations in the local microwave properties of the sample) and the topography of the sample. A simple circuit model (for a sample with dominant ohmic losses) includes the probe-to-sample capacitance ($C_x$) and the sheet resistance ($R_x$) of the sample (Fig. 1 inset). This particular microscope has a lateral spatial-resolution in $C_x$ of at least 2.5 nm.[23]

The interaction of the tip and sample for the STM-assisted NSMM is non-trivial. The geometry of the tip can have a significant effect on the quality of the signals from the integrated microscope. In Fig. 2, we show optical and SEM micrographs of a set of commercially available tips used in our experiment. The tips reported here are from two companies, Materials Analytical Services[28] (MAS) and Advanced Probing Systems[29] (the commercial name for each tip starts with "WRAP"). The MAS tip is Pt/Ir while the WRAP tips are Ag-coated W tips. The first two columns of the figure present optical micrographs (second column is at higher magnification) of the tips and the third column shows scanning electron microscope (SEM) images of the tips. From the optical pictures, we make the observation that all of the tips have an overall conical geometry. In the case of the MAS tip[28], it actually shows multiple cones embedded one into another. However, the very end of the tip can be envisioned as a sphere embedded at the end of a cone (shown in the SEM column of Fig. 2).

The estimated radius of the sphere ($r_{sphere}$), for different tips is given in Table 1 and it varies from 100 nm to 8μm. The larger values of $r_{sphere}$ clearly help in getting larger $\Delta f$ contrast over a given sample. The reported frequency shift contrast ($\overline{\overline{\Delta f}}$) in Table 1 is the frequency change measured between the tunneling height and 2 μm away from the various samples. For example, above the thin film of gold on glass, the $\overline{\overline{\Delta f}}$ is 35 kHz with the tip MAS tip ($r_{sphere}$ of 100 nm) and above the same film the $\overline{\overline{\Delta f}}$ is 1275 kHz with the WRAP30R tip[29] ($r_{sphere}$ of 8 μm). Similar trends are seen above thin film of gold on mica and bulk copper samples. We find that this $\overline{\overline{\Delta f}}$ increases roughly exponentially (for a given sample) with increasing $r_{sphere}$ of different probes.[6] Such results argue that good NSMM signal contrast is associated with tips with large $r_{sphere}$. However, such tips are not necessarily good for STM.

Fig. 3 shows the STM topography imaging done with three different tips for a standard Nickel CD master sample used for calibration purposes of STMs and AFMs.[34] The sample is a 0.3 mm thick Nickel disc, which has the diameter of 6.3 mm. The features are 100 nm in topography and the pitch is 740 nm. As can be seen in Fig. 3, the topography of 100 nm is well re-produced by all three tips, however, the pitch of 740 nm is not re-produced well by WRAP30D tip (compare to the AFM image in Fig. 3). The WRAP178 reproduces the pitch well, however, it makes the boundaries of the features very 'ragged'. The WRAP082 reproduces both the topography and the pitch very well. The WRAP30D has $r_{sphere}$ of 5 μm and at the pitch of the Nickel sample, it probably produces multiple tunneling sites during a scan which distort the size of the features laterally. On the other hand, the WRAP082 ($r_{sphere}$ = 550 nm) and WRAP178 ($r_{sphere}$ = 130 nm) reproduce the expected images. The 'raggedness' of the features with WRAP178 may be due to mechanical vibrations present in the lateral direction, associated with the thin geometry of the tip (see Fig. 2) and its unusually long projection from the capillary tube. Besides, the tip-to-sample capacitance ($C_x$) is very small with WRAP178, so overall the WRAP082 tip makes a good compromise for both microscopes.

## III. Model of Tip-Sample Interaction:

Since the nominal tip-to-sample separation during scanning is 1 nm, contributions from evanescent waves must be part of the theoretical model to calculate the measured quantities i.e., Δf and Q of the NSMM. For the transmission line resonator-based NSMM, transmission line models[5,23] and 2D-electrostatic numerical models[6] have already been discussed in the literature. In this paper, we discuss a new full-wave model, which gives insight into the high spatial-resolution of this microscope and includes contributions from evanescent waves. We discuss this quantitative model and use it to calculate the Δf and Q of the resonator when a sample is present.

The first step to understand the tip-to-sample interaction is to use the model of a conducting sphere above an infinite conducting plane (sphere-above-the-plane as shown in Fig. 4). The geometry of the model is similar to others discussed before[26,27], however, this time, in order to calculate the fields, the sphere will be replaced by an oscillating infinitesimal vertical dipole[31,32] of electric moment $I_0\ell$, where $I_0$ is the dipole current and $\ell$ is its length. The strength of this model lies in providing the complete electromagnetic field in the region[31,32] between the tip and the sample. The essential length scales in the problem are the sphere radius $R_0$ and height h of the sphere above the plane. The model is motivated by the presence of the embedded sphere at the very end of the tip (see SEM column of Fig. 2).

The geometry of the model is shown in Fig. 4, and in Figs. 4(a) and 4(b) we schematically show the electric $\vec{E}$ and magnetic $\vec{B}$ fields, respectively (a cylindrical coordinate system is used with z being positive into the sample). The electric field at the sample surface is in the $\hat{z}$ and $\hat{\rho}$ directions and magnetic field is in the $\hat{\phi}$ direction. The probe is now modeled as a sphere of radius $R_0$ and this sphere will be replaced by an infinitesimal vertical dipole, located at the center of the sphere, as shown in Fig. 4(d). For our experiment, the vertical dipole will always be a distance $h+R_0$ away from the sample surface (Fig. 4(c)).

We follow the convention of the vertical radiating dipole model[31,32], where the region of the material (sample region) is labeled 1 and the dipole (probe) is in region 2. This model considers an infinitesimal oscillating vertical electric dipole a distance h above the planar interface between regions 1 and 2. Region 1 is described by wave vector $k_1$, dielectric constant $\varepsilon_1$ and conductivity $\sigma_1$, whereas region 2 is described by wave vector $k_2$, dielectric constant $\varepsilon_2 = \varepsilon_0$ and conductivity $\sigma_2=0$. In order to calculate the microscope properties $\Delta f$ and Q, the stored energy and the power dissipated in the sample must be calculated. The dissipated power ($P_{dissipated}$) and stored energy ($U_{sample}$) inside the sample are calculated using the integrated Poynting vector $I_{poynting} = \frac{1}{2\mu_0} \oint_S (\vec{E} \times \vec{B}^*) \cdot d\vec{a}$, where the integral is performed over the planar interfaces between media 1 and 2. The real part of the $I_{poynting}$ gives the dissipated power ($P_{dissipated}$) and the imaginary part gives $2\omega$ times the stored energy ($U_{sample}$) in the sample.[40]

The plot of dissipated power versus height calculated from this vertical radiating dipole model for several different samples is shown in Fig. 5. The important features to note are that the dissipated power increases two ways: one, as the vertical radiating dipole is brought near the sample of given skin depth, $\delta$ (which is a measure of conductivity of the material) and two, at a fixed height when the resistivity of the sample increases (increasing δ). The saturation that is seen in the curves as h goes to 1 nm is due to the fact that actual height of the vertical dipole is h+$R_0$ (Fig. 4(c)). This is very similar to the saturation in the frequency shift signal that we reported earlier[6], which shows that the near-field signal is sensitive to another length scale besides the height of the probe above the sample. We discussed earlier[6] that this additional length scale can be explained by adding a small irregularity at that end of the probe (experimentally a "particle" sticking to the probe) which illuminates the sample. Here this additional length scale comes from the sphere which hosts the vertical dipole (Fig. 4(d)). The change in Quality Factor (Q) of the resonator due to the sample is calculated from this dissipated power as

$$\frac{1}{Q`} - \frac{1}{Q_0} = \frac{P_{dissipated}}{\omega U_{resonator}}, \quad (1)$$

where $U_{resonator}$ is the stored energy in the resonator, the sample, and the coupling fields between them and $Q_0$ is the resonator quality factor in the absence of a sample. Equation (1) implicitly assumes that there is no change is dissipated power in the probe and microscope as the height h is changed.

The stored energy in the sample ($U_{sample}$) ranges roughly from $10^{-24}$ to $10^{-17}$ J and is plotted as dashed curves in Fig. 5 (right y-axis). We see that the stored energy in the sample in all cases is increasing as the height of the vertical dipole is reduced. This is expected from the field equations[31,32] which shows that the fields increase as we reduce the height of the vertical dipole above the sample. Again, the saturation happens at h~$R_0$, since the sphere of radius $R_0$ hosts the vertical dipole. The energy stored in the sample is the lowest in the case of Copper, as it is the closest to the case of a perfect dielectric of the three samples shown. The frequency shift of the microscope due to a metallic sample (Δf) can be calculated as the ratio of stored energy in the sample and between the tip and the sample, compared to the total energy stored in the resonator, given by the equation

$$\frac{\Delta f}{f} = -\frac{U_{sample} + \frac{1}{2}C_x V^2}{U_{resonator}} \quad , \quad (2)$$

where V is the tip-to-sample potential difference. In the next section we compare the model results with data.

## IV. Experimental Results and Comparison to Model:

We find from the calculation that the energy stored inside the samples considered here (Copper and Silicon) is $\sim 10^{-24}$ to $10^{-17}$ J, which is much smaller than the energy stored due to capacitance between tip and a conducting sample ($\frac{1}{2}C_x V^2 \sim 0.3 \times 10^{-15}$ J). Hence, the major contribution to Δf comes from the capacitance term (energy stored in electric fields between the tip and the sample), and this has been discussed earlier.[23]

However, for the Q, the results are non-trivial and in Fig. 6, we plot the calculated resonator quality factor (Q) as a function of height and compare it to the data taken over a bulk uniform Silicon wafer and bulk Copper. The Silicon wafer is n-type with nominal resistivity $\rho = 20$ Ω.cm at room temperature, and thickness of 550 μm. The calculation has been performed for the cases of Copper (δ=0.75 μm) and Silicon (δ=2570 μm) at 7.67 GHz. The $\hat{Q}$ (resonator quality factor with sample present) is calculated using (1), where $Q_0$ is the experimentally determined quality factor when no sample is present. The data on a bulk silicon wafer is in good agreement with the model calculation, where the two fit parameters $I_0\ell = 0.53 \times 10^{-10}$ A.m and $R_0 = 76$ nm. The calculations predict that the drop in Q upon approaching the sample is higher for larger sample skin depth $\delta$. A low loss bulk copper case (data and fit to $\delta = 0.75$ μm and the rest of the parameters are the same as for bulk Silicon) is also shown. The data on a bulk copper sample demonstrates the fact that we do not see any significant Q(h) contrast in a low loss bulk sample. Hence, the Silicon and Copper data together demonstrate that the quality factor is sensitive to a materials property of the sample.

We have also performed experiments on thin films, and in this case the field and current distributions in the sample can be substantially different from the bulk case (hence the model discussed above is not applicable). Here we present data for thin films in the high and low limits of sheet resistance. An example of the image data on a laterally inhomogeneous colossal magneto-resistive (CMR) $La_{0.67}Ca_{0.33}MnO_3$ thin film[36,37] on $LaAlO_3$ substrate is shown in Fig. 7. The images are 530 nm square and all three images were acquired simultaneously. The STM was operating in constant tunnel current mode, with a bias of 1 volt on the sample and tunnel current set point of 1 nA. Fig. 7(a) shows the STM topography of the sample and Figs. 7(b) and 7(c) show resonator quality factor (Q) and frequency shift ($\Delta f$), respectively. The nominal sheet resistance ($R_x$) of the CMR thin film is ~1kΩ/□, which is a high-$R_x$ film as compared to a low-$R_x$ ($R_x$ ~ 0.1 Ω/□) gold on mica thin film shown in Fig. 8.

In Fig. 8, we show simultaneously acquired images of STM, Q and $\Delta f$ on a thin film of gold on mica substrate. This time only the $\Delta f$ image (Fig. 8(c)) shows correlation with the STM topography image, while the $V_{2f}$ image in Fig. 8(b) ($V_{2f} \sim Q$) does not show any significant contrast. These experiments raise two interesting questions: why is it that the spatial-resolution of the NSMM is apparently no worse than the STM itself[23], despite the fact that the probe illuminates an area at least on the order of $r_{sphere}^2$ with strong fields; and the second is why there is Q contrast only for the CMR thin film and not the gold on mica thin film?

In order to understand these observations, we need to refer to the simple model of tip-to-sample interaction (shown in the inset of Fig. 1) which models losses in the sample as a resistance $R_x$ in series with the tip-to-sample capacitance $C_x$. When the STM is in distance-following mode, the probe follows the topography of the sample. During distance following, the NSMM sees changes in $C_x$, which to first order approximation[6,23] dominates the frequency shift $\Delta f$, resulting in a strong correlation between $\Delta f$ and STM topography for both samples. The spatial-resolution of the NSMM signal is much better than the probing length-scale $r_{sphere}$ due to STM distance-following control[23], since it maintains a nominal height of 1 nm above the sample.

However, the microscope Q is sensitive to the losses of the sample (Fig. 6), and during scanning it gives a measure of the local loss. Hence, we expect to see local loss variations in the high sheet resistance CMR thin film while we do not expect any local loss contrast in the low-loss gold on mica thin film. There is already a good discussion in the literature[25] of the experimental dependence of microscope Q on $R_x$, which predicts that for a low $R_x$ ($\omega C_x R_x \ll 1$) film the Q contrast is small compared to high $R_x$ films ($\omega C_x R_x$ approaching 1).

# V. Future Directions:

*Theoretical models:* We have used the model of an electrically small radiating dipole antenna near an interface between two semi-infinite media[31,32] to describe contrast from a near-field microwave microscope. This problem offers a full-wave self-consistent solution to Maxwell's equations (and it includes evanescent wave contributions) in both media. The problem can be extended to multi-layered structures, and is already partially solved for a three layer structure.[39] However questions have arisen about the near-field structure in the solution for King's dipole-above-the-plane model for lossy dielectric samples.[38] Further investigation of the approximation[31,39] made to derive a closed-form expression for the electromagnetic field is required.

We have found from the field equations[31,32] that reducing $R_0$ helps in increasing the magnitude and the spatial-confinement of the electromagnetic fields. Hence, for future work it is important to find out how the geometry of the probe relates to the vertical dipole strength parameter $I_0\ell$, so that we can design probes with enhanced loss sensitivity and decreased confinement region of the fields.

*Experiments:* It is desirable in NSMM experiments that high spatial-resolution techniques be extended to dielectric (insulating) samples as well.[33, 35] For such samples, STM is not a viable option for probe-positioning. The atomic force microscope (AFM) is better suited for distance-following on dielectric samples and AFM probes are also much more robust compared to STM tips, although quantitative imaging is difficult to achieve.

## Acknowledgements:

We thank Bobby Moreland at Neocera for his help with the SEM imaging, and Amlan Biswas and R. L. Greene for preparing the CMR thin films. This work has been supported by an NSF Instrumentation for Materials Research Grant DMR-9802756, the University of Maryland/Rutgers NSF-MRSEC and its Near Field Microwave Microscope Shared Experimental Facility under Grant number DMR-00-80008, NSF/GOALI DMR-0201261, the Maryland Industrial Partnerships Program 990517-7709, the Maryland Center for Superconductivity Research and by a Neocera subcontract on NIST-ATP# 70NANB2H3005.

**Figure Captions:**

**Figure 1: Schematic of the transmission line resonator based STM-assisted NSMM. A bias-Tee is included to couple the STM feedback circuit to the center conductor of the resonator. The FFC keeps the source locked onto one resonant frequency of the resonator. Three independent quantities are measured; STM topography, resonator quality factor (Q) and the frequency shift (Δf), shown in boxes.**

**Figure 2:** (Color Online) Optical and SEM micro-graphs of different commercially available STM tips. Optically, the conical nature of the tips is clear (the first two columns). The optical 1 column is at a magnification X 10 (of objective lens) and the optical 2 column is at magnification of X 40. The third column shows SEM micro-graphs of the same tips to clarify the embedded sphere at the end of each conical tip.

**Figure 3:** (Color Online) STM topography images of a CD master sample from three different tips: WRAP30D, WRAP082 and WRAP178. The WRAP082 is the best compromise for both STM and NSMM. An AFM image of the CD master is added for reference, where the manufacturer claims the features are 100 nm high.

**Figure 4:** Schematic diagram to show the directions of a) electric and b) magnetic fields due to the sphere. Note that the z-direction is positive into the sample. Part c) shows the effective height of the dipole above the sample while in d) the vertical dipole replaces the sphere.

**Figure 5:** (Color Online) Calculated dissipated power (solid curves left y-axis) over different samples as a function of height, as calculated from the vertical radiating dipole model. The calculation is performed for a frequency of 7.67 GHz, $I_0 \ell = 0.53 \times 10^{-10}$ A.m, $R_0$ of 76 nm and for three samples characterized by their skin-depth (conductivity). Stored energy in the sample (dashed curves right y-axis) versus height of the vertical dipole above the same samples. Note that the saturation starts to appear when $h \sim R_0$.

**Figure 6: (Color Online) The Q˙ (resonator quality factor with the sample present) is calculated using** $\frac{1}{Q\grave{}} - \frac{1}{Q_0} = \frac{P_{dissipated}}{\omega U_{resonator}}$ **and ω, $Q_0$ (experimentally determined quality factor when no sample is present), $U_{resonator}$ (stored energy in the resonator is $0.6 \times 10^{-12}$ J at an input power of 1mW) and $P_{dissipated}$ in the sample are all known quantities. The data on a bulk silicon wafer is in good agreement with the model calculation, where the model parameters are frequency of 7.67 GHz, δ =2570 µm, $I_0\ell = 0.53 \times 10^{-10}$ A.m and $R_0$ = 76nm. Comparison of calculated Q with the data performed with the WRAP082 tip. A low loss bulk copper case (data and fit with δ = 0.75 µm and all other parameters the same as bulk Silicon) is also shown.**

**Figure 7:** (Color Online) **Simultaneously acquired Q and Δf images for $La_{0.67}Ca_{0.33}MnO_3$ thin film on $LaAlO_3$ substrate. The experiment was performed at T = 240 K (below the Curie temperature for this particular sample) using a WaveTek 907 source at a frequency of 7.67 GHz. The $V_{bias}$ for STM was 1 volt and tunnel current set point was 1 nA. The experiments were performed with a Pt/Ir etched tip with geometry similar to WRAP082. Note that the Q and Δf images show contrast on the same scale as variations in the STM topography image.**

**Figure 8:** (Color Online) **Simultaneous imaging of a gold on mica thin film sample. The bias for the sample is 0.1 volts and tunnel current set point is 1 nA. This is a room temperature experiment performed at 7.48 GHz with a WaveTek 904 source. The $V_{2f}$ presented here is proportional to the Q (Q ≅ 384) of the resonator, and the image is**

**noisy because no changes in losses are detected (the changes in Q are on the order of 0.1). The experiments were performed with Pt/Ir etched tips with geometry similar to WRAP082.**

# References:

[1] E. A. Ash and G. Nicholls, Nature **237**, 510 (1972)

[2] D. W. Pohl, W. Denk, and M. Lanz, Appl. Phys. Lett. **44**, 651 (1984).

[3] E. Betzig, M. Issacson and A. Lewis, Appl. Phys. Lett. **51**, 2088 (1987).

[4] S. M. Anlage, D. E. Steinhauer, B. J. Feenstra, C. P. Vlahacos and F. C. Wellstood, in *Microwave Superconductivity*, ed. By H. Weinstock and M. Nisenoff, (Klumwer, Amsterdam, 2001), p.239.

[5] B. T. Rosner, D. W. van der Weide, Rev. Sci. Instrum. **73**, 2505 (2002).

[6] A. Imtiaz, M. Pollak, S. M. Anlage, J. D. Barry and J. Melngailis, J. of Appl. Phys., **97**, 044302 (2005).

[7] A. Kramer, F. Keilmann, B. Knoll and R. Guckenberger, Micron, **27**, 413 (1996).

[8] F. Keilmann, D. W. van der Weide, T. Eickelkamp, R. Merz and D. Stockle, Optics Commun., **129**, 15 (1996).

[9] S. J. Stranick and P. S. Weiss, Rev. Sci. Instrum. **64**, p1232 (1993). Erratum: Rev. Sci. Instrum. **64**, 2039 (1993).

[10] S. J. Stranick, M. M. Kamna and P. S. Weiss, Rev. Sci Insturm. **65**, 3211 (1994).

[11] S. J. Stranick and P. S. Weiss, Rev. Sci. Instrum. **65**, 918 (1994).

[12] B. Knoll, F. Keilmann, A. Kramer and R. Guckenberger, Appl. Phys. Lett. **70,** 2667 (1997).

[13] B. Michel, W. Mizutani, R. Schierle, A. Jarosch, W. Knop, H. Benedickter, W. Bachold and H. Rohrer, Rev. Sci. Instrum. **63**, 4080 (1992).

[14] W. Mizutani, B. Michel, R. Schierle, H. Wolf and H. Rohrer, Appl. Phys. Lett. **63**, 147 (1993).

[15] Greg P. Kochanski, Phys. Rev. Lett. **62**, 2285 (1989).

[16] W. Seifert, E. Gerner, M. Stachel and K. Dransfeld, Ultramicroscopy **42-44**, 379 (1992).

[17] L. A. Bumm and P. S. Weiss, Rev. Sci. Instrum. **66**, 4140 (1995).

[18] W. Kreiger, T. Suzuki and M. Völcker, Phys. Rev. B **41**, 10229 (1990).

[19] M. Völcker, W. Krieger, T. Suzuki and H. Walther, J. Vac. Sci. Technol. B **9**, 541 (1991).

[20] M. Völcker, W. Krieger and H. Walther, Phys. Rev. Lett. **66**, 1717 (1991).

[21] M. Völcker, W. Krieger, and H. Walther, J. Appl. Phys. **74**, 5426 (1993).

[22] G. Nunes Jr. and M. R. Freeman, Science **262**, 1029 (1993).

[23] A. Imtiaz and S. M. Anlage, Ultramicroscopy, **94,** 209 (2003).

[24] D. E. Steinhauer, C. P. Vlahacos, S. K. Dutta, F.C. Wellstood and Steven M. Anlage, Appl. Phys. Lett. **71**, 1736 (1997).

[25] D. E. Steinhauer, C. P. Vlahacos, S. K. Dutta, B. J. Feenstra, F.C. Wellstood and Steven M. Anlage, Appl. Phys. Lett. **72**, 861 (1998).

[26] Chen Gao, Fred Duewer and X.-D.Xiang, Appl. Phys. Lett. **75**, 3005 (1999). Erratum; Appl. Phys. Lett. **76**, 656 (2000).

[27] Xiao-Dong Xiang, Chen Gao, Peter. G. Schultz and Tao Wei, *"Scanning Evanescent Electro-Magnetic Microscope"* US patent# 6532806 B1, (2003).

[28] Materials Analytical Services; www.mastest.com;  2418 Blue Ridge Road, Suite 105, Raleigh, NC 27607 USA.

[29] Advanced Probing Systems; www.Advancedprobing.com; P.O Box 17548, Boulder, CO 80308 USA.

[30] Advanced Surface Microscopy, Inc.; www.asmicro.com; 3250 N. Post Rd., Suite 120, Indianapolis, IN 46226 USA.

[31] Ronold W. P. King, Radio Science, **25**, 149 (1990).

[32] Ronold W. P. King, Margaret Owens and Tai Tsun Wu *Lateral Electromagnetic waves: Theory and application to communications, geophysical exploration, and remote sensing*, Springer-Verlag, New York (1992).

[33] "*International Technology Roadmap for Semiconductors: Metrology*" by Semiconductor Industry Association (2004).

[34] This sample is available from Advanced Surface Microscopy[30] (commercial name of sample is HD-750).

[35] Takeshi Morita and Yasuo Cho, Appl. Phys. Lett. **84**, 257 (2004).

[36] Amlan Biswas, M. Rajeswari, R. C. Srivastava, Y. H. Li, T. Venkatesan, and R. L. Greene, Phys. Rev. B **61**, 9665 (2000).

[37] Amlan Biswas, M. Rajeswari, R. C. Srivastava, T. Venkatesan, and R. L. Greene, Phys. Rev. B **63**, 184424 (2001).

[38] R. E. Collin, IEEE Trans. on Anten. and Prop. **52**, 3133 (2004).

[39] Ronold W. P. King and Sheldon S. Sandler, Radio Science, **29**, 97 (1994).

[40] John D. Jackson, *Classical Electrodynamics*, 3rd Edition, Wiley, New York, 1999, p. 265.

Table 1: Summary of results from the tip study used for the STM-assisted NSMM. The $\overline{\overline{\Delta f}}$ is defined as contrast in Δf from tunneling heights above the sample to 2 μm (maximum retraction height of the z-piezo of the STM) above the sample, i.e. $\overline{\overline{\Delta f}} = \Delta f(2\mu m) - \Delta f(1 nm)$.

| Company Name | Brief Description of tip | Radius of Curvature ($r_{sphere}$) | $\overline{\overline{\Delta f}}$ contrast over bulk copper | $\overline{\overline{\Delta f}}$ contrast above gold on glass film | $\overline{\overline{\Delta f}}$ contrast above gold on mica film | STM Quality |
|---|---|---|---|---|---|---|
| *Materials Analytic Services (MAS)* | *Platinum-Iridium etched tip* | 100 nm | 65 kHz | 35 kHz | 21 kHz | Good (gold on mica) |
| *Advance Probing Systems (APS)* | *Ag coated W etched tip WRAP178* | 130 nm | 45 kHz | 40 kHz | 55 kHz | Underestimates the feature size (Ni CD-master sample) |
| *Advance Probing Systems (APS)* | *Ag coated W etched tip WRAP082* | 550 nm | 85 kHz | 135 kHz | 40 kHz | Good (many samples) |
| *Advance Probing Systems (APS)* | *Ag coated W etched tip WRAP30D* | 5 μm | 300 kHz | 1150 kHz | 300 kHz | Doubles feature size (Ni CD-master sample) |
| *Advance Probing Systems (APS)* | *Ag coated W etched tip WRAP30R* | 8 μm | 1000 kHz | 1275 kHz | 1100 kHz | Doubles feature size (Ni CD-master sample) |

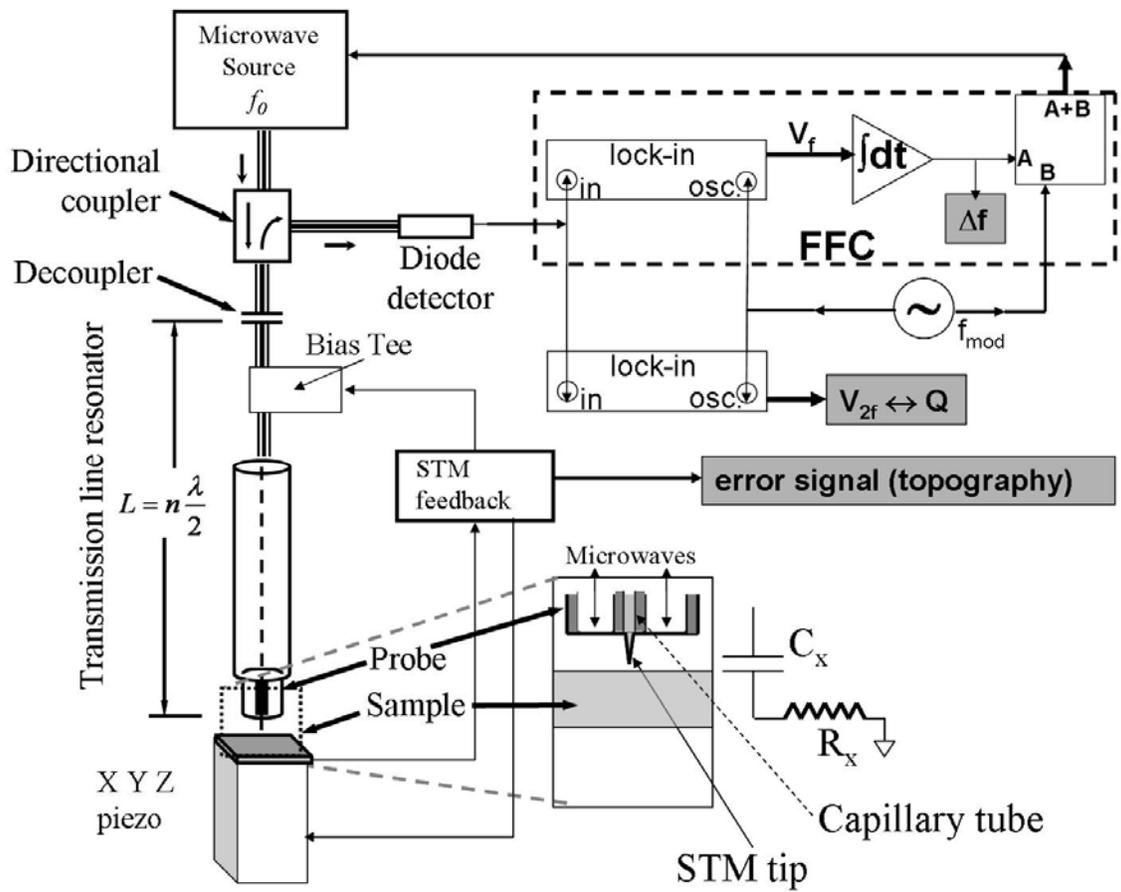

Fig. 1: Atif Imtiaz and Steven M. Anlage

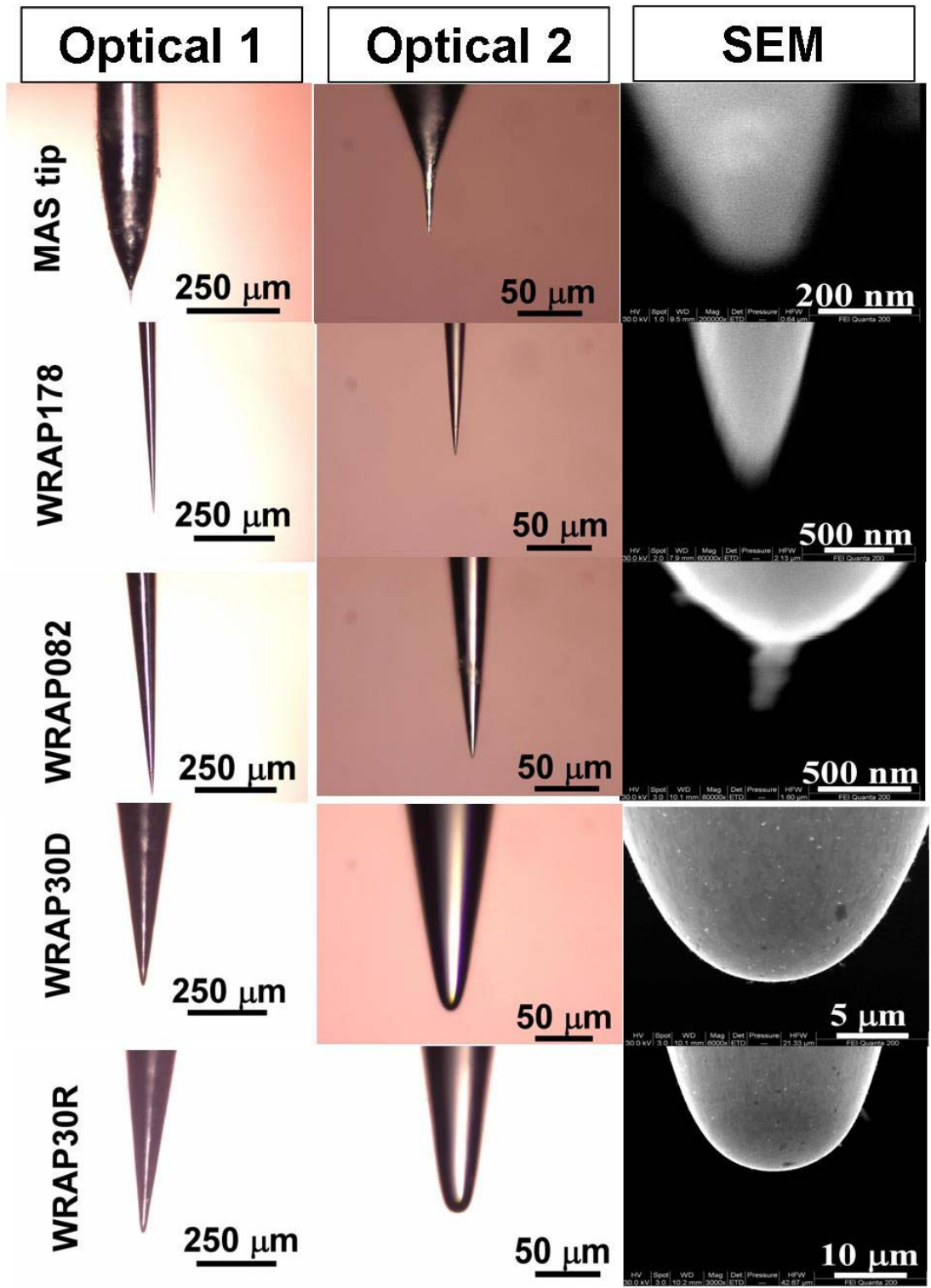

Fig. 2: Atif Imtiaz and Steven M. Anlage

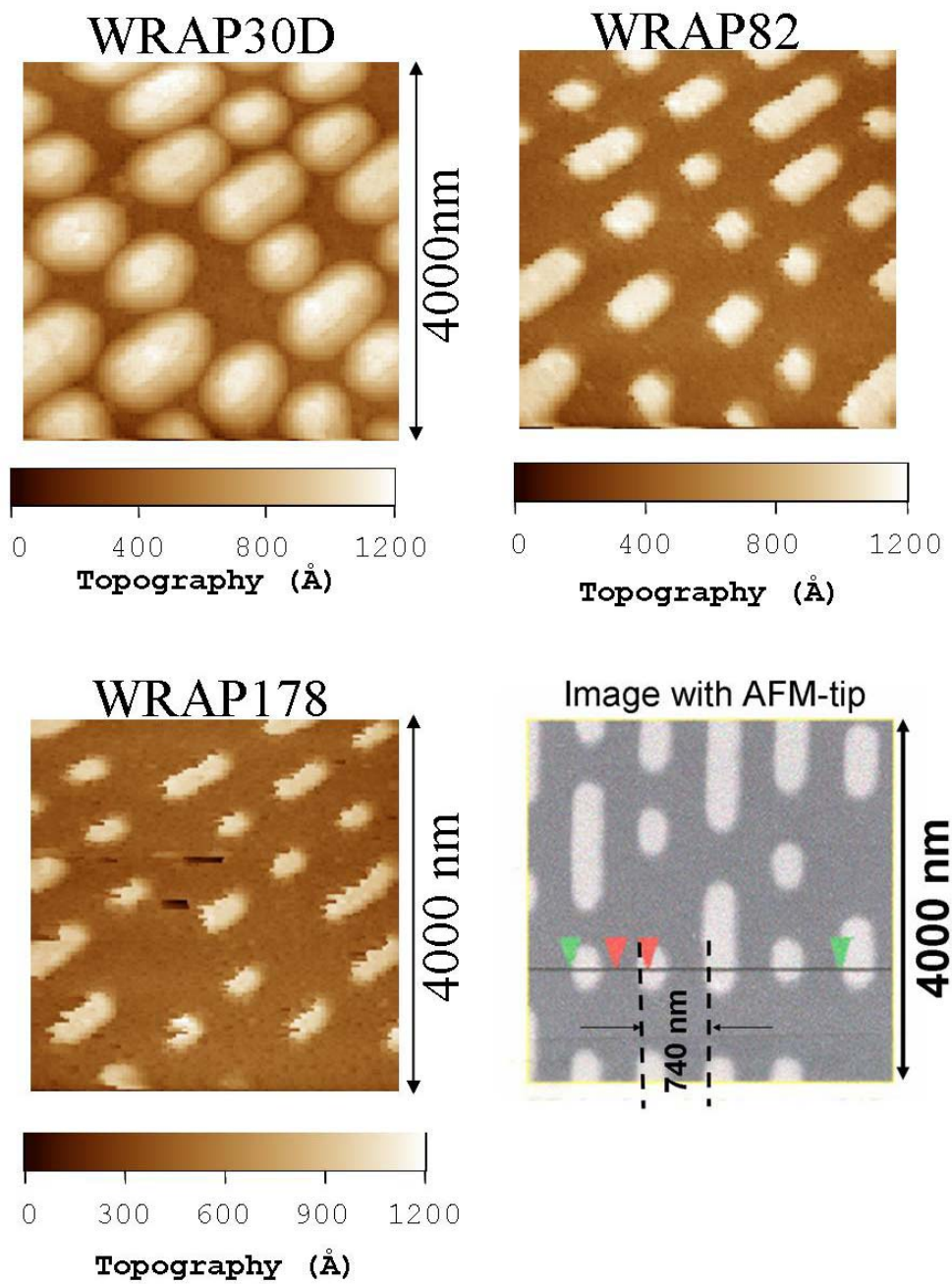

Fig. 3: Atif Imtiaz and Steven M. Anlage

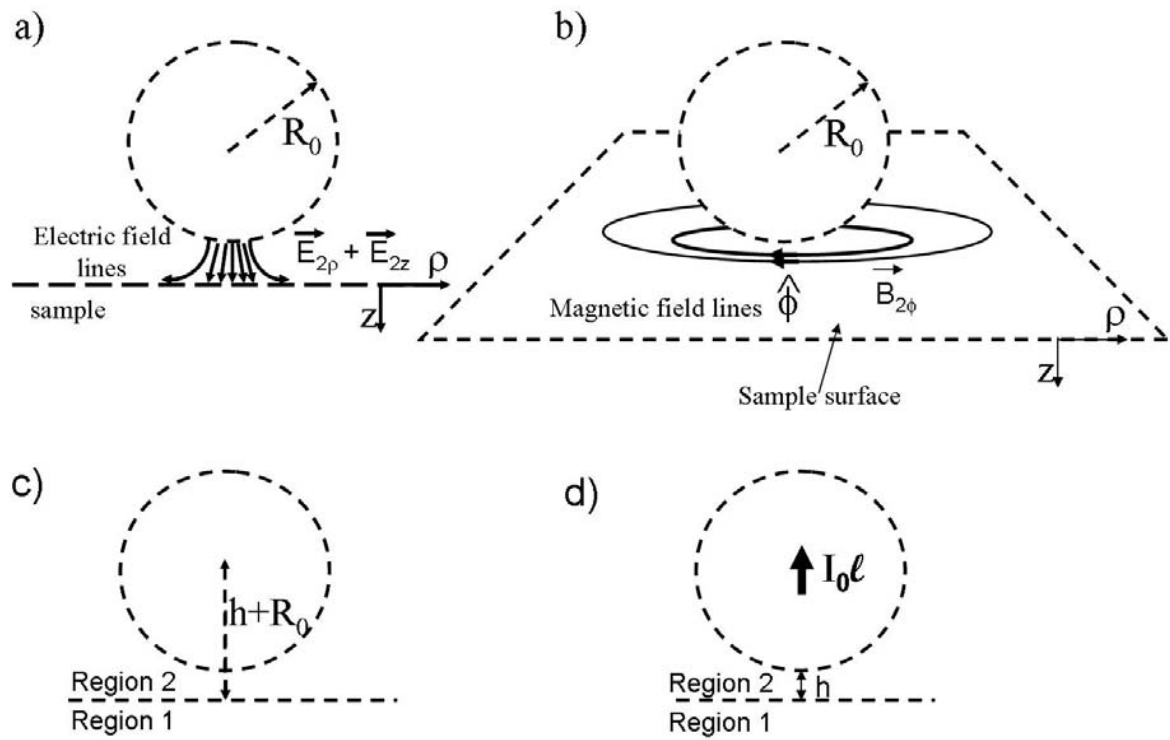

Fig. 4: Atif Imtiaz and Steven M. Anlage

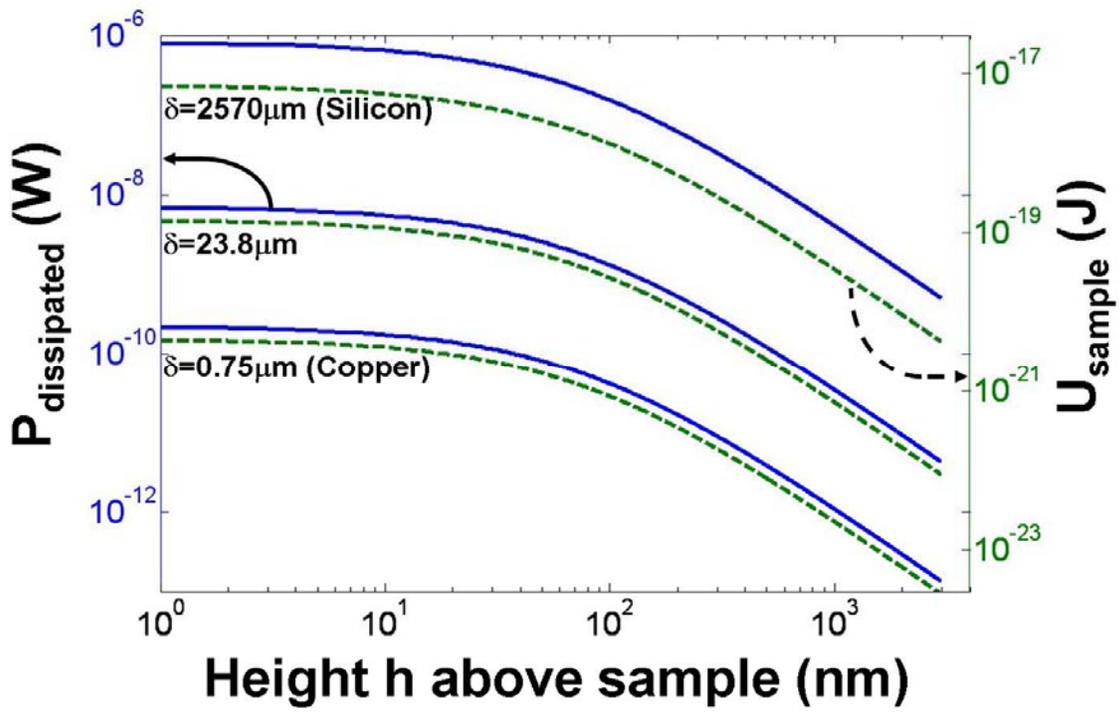

Fig. 5: Atif Imtiaz and Steven M. Anlage

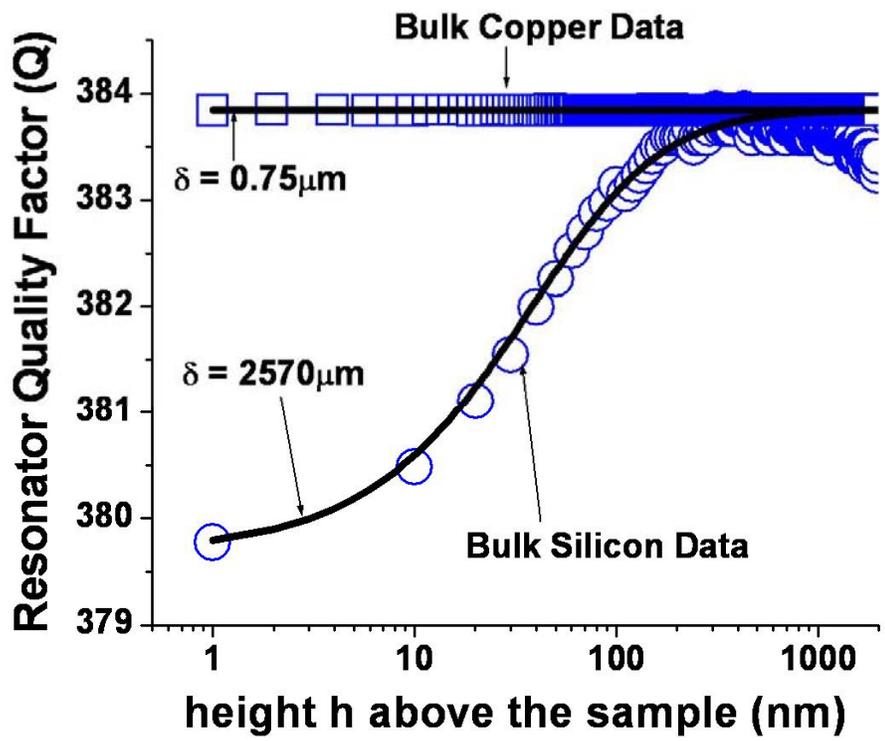

Fig. 6: Atif Imtiaz and Steven M. Anlage

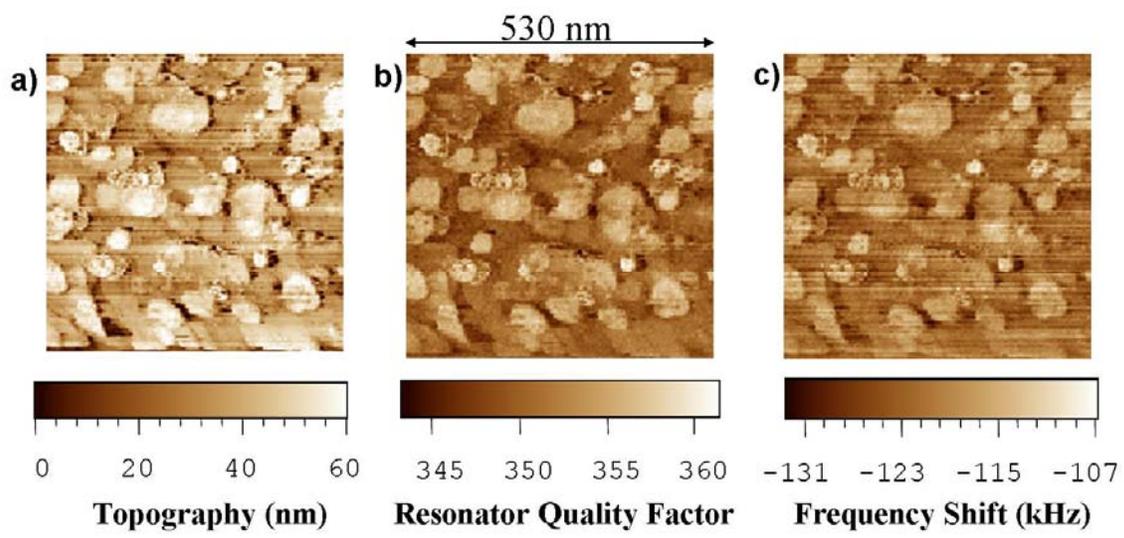

Fig. 7: Atif Imtiaz and Steven M. Anlage

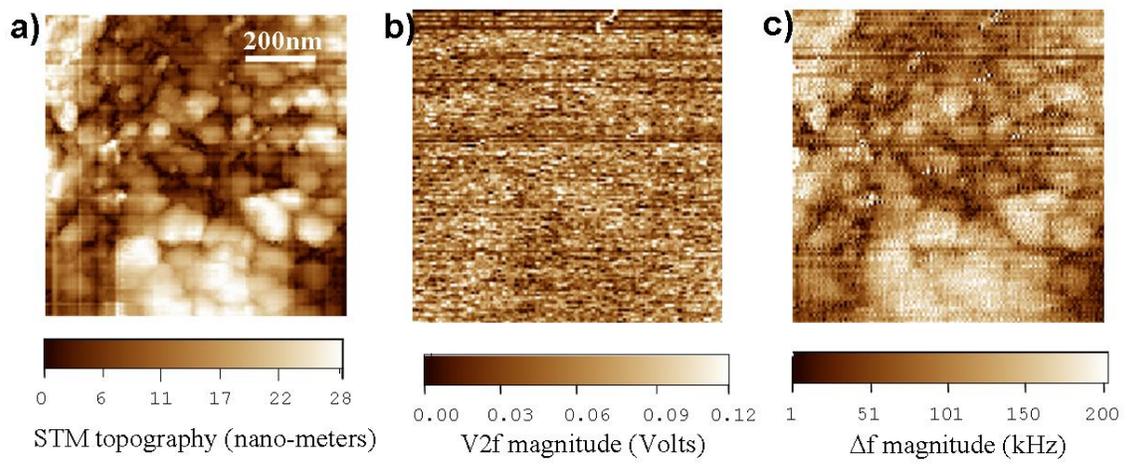

Fig. 8: Atif Imtiaz and Steven M. Anlage